\documentclass[fleqn,usenatbib]{mnras}
\usepackage{graphicx}	
\usepackage{amsmath}	
\usepackage{amssymb}	

\title[GRB 141221A]{GRB 141221A: gone is the wind}

\author[O. Bardho et al.]{O.~Bardho$^{1, 2}$, B.~Gendre$^{3, 4, 1}$, A.~Rossi$^{2}$,  L.~Amati$^{2}$, J.~Haislip$^5$, A.~Klotz$^{6}$
\newauthor E.~Palazzi$^{2}$, D.~Reichart$^5$, A.~S.~Trotter$^{5, 7}$,  M.~Bo\"{e}r$^{1}$\thanks{Corresponding author: michel.boer@unice.fr}\\
$^1$ARTEMIS, CNRS UMR 5270, Universit\'{e} C\^{o}te d'Azur, Observatoire de la C\^{o}te d'Azur, \\
\, boulevard de l'Observatoire, CS 34229, F-06304 Nice Cedex 04, France\\
$^2$IASF Bologna-INAF, via P. Gobetti 101, Bologna, Italy\\
$^3$Etelman Observatory, St Thomas, Virgin Islands, USA\\
$^4$University of the Virgin Islands, 2 John Brewer's Bay Road, 00802 st Thomas, Virgin Islands, USA\\
$^5$Department of Physics and Astronomy, University of North Carolina at Chapel Hill, Campus Box 3255, Chapel Hill, NC 27599, USA\\
$^6$IRAP, CNRS, Universit\'e de Toulouse, 9 avenue du colonel Roche, 31028 Toulouse Cedex 4, France\\
$^7$Department of Physics, NC A\&T State University, 1601 E. Market St, Greensboro, NC 27411, USA\\ }

\begin{document}

\date{Received-- Accepted}

\pagerange{\pageref{firstpage}--\pageref{lastpage}} \pubyear{2016}

\maketitle

\label{firstpage}

\begin{abstract}
GRB 141221A was observed from infrared to soft gamma-ray bands. Here, we investigate its properties, in light of the standard model. We find that the optical light curve of the afterglow of this burst presents an unusual steep/quick rise. The broad band spectral energy distribution taken near the maximum of the optical emission presents either a thermal component or a spectral break. In the former case, the properties of the afterglow are then very unusual, but could explain the lack of apparent jet breaks in the Swift light curves. In the latter case, the afterglow properties of this burst are more usual, and we can see in the light curves the passing through of the injection and cooling frequencies within the optical bands, not masked by a reverse shock. This model also excludes the presence of a stellar wind, challenging either the stellar progenitor properties, or the very stellar nature of the progenitor itself. In all cases, this burst may be a part of a Rosetta stone that could help to explain some of the most striking features discovered by Swift during the last ten years.

\end{abstract}


\section{Introduction}
Since the launch of the {\em Swift} satellite in 2004 \citep{gehrels_04}, hundreds of gamma-ray bursts \citep[GRB, see][for a review]{kum15} have been detected, localized and followed both on-board and by telescopes on the ground. This led to a very large sample of events presenting virtually all possible aspects of the standard model \citep[see][for a complete description of the model]{ree92, mes97, pan98}. Several events have been followed in optical with rapid robotic telescopes while the prompt emission was still active or recently concluded, and in a fair number of cases a rising behavior has been observed in this band \citep[see for instance][]{gen12}.

This rise of the optical wavelength emission can be understood in two different ways: either it is the initial part of the forward shock, which can be observed until the injection frequency $\nu_m$ crosses the observational band, or we see the signature of the reverse shock \citep[e.g.][]{sar99}. Both phenomena can be interleaved, complicating the analysis.

GRB 141221A is one of these "optically rising" bursts. It was detected by {\em Swift} at 08:07:10 UT (hereafter $T_{0}$) on December 21, 2014 \citep{b1}. The duration of the burst, while not exceptional \citep[$T_{90}$ = 36.9 $\pm$ 4.0 s,][]{b2}, allowed the TAROT and Skynet robotic observatories to start the observation while the prompt emission was still active. While in other cases the rise was smooth and not extreme, in this case the optical emission increased very quickly and presented other features usually not seen; the purpose of this work is to investigate those features.

In Section \ref{sec_data} we present the data for this event. We explain the data reduction in Section \ref{sec_red}, and present the spectral and temporal analyses in Section \ref{sec_ana}. We then discuss our results in Section \ref{sec_discu}, before concluding.

In the remainder of this paper, all errors are quoted at the 90\% confidence level (except when otherwise stated), and we use a flat $\Lambda$CDM model for the Universe, with $H_{0}$ = 70 km s$^{-1}$ Mpc$^{-1}$, $\Omega_{m}$ = 0.3 and $\Omega_{\Lambda}$ = 0.73. We will use the standard notation $F_{\nu}\propto t^{-\alpha} \nu^{-\beta}$.

\section{Observations}
\label{sec_data}
\subsection{High Energy data}
{\bfseries{{\em Swift}-BAT and Fermi GBM}}: GRB 141221A triggered both instruments \citep{b2, b12}, at nearly the same time (08:07:10 UT for Swift, 08:07:11.22 UT for Fermi). The recorded duration is, however, longer in the BAT compared to GBM (23.8 s), as one can expect from the larger effective area (and hence better sensitivity) of BAT/{\em Swift}.

{\bfseries{{\em Swift}-XRT}}: The XRT observed the burst position between $T_{0}$ + 64 s and $T_{0}$ + 34.9 ks \citep{b7, b10}, mostly in PC mode. The afterglow was clearly detected in X-rays.

\subsection{Optical and infrared data}
Table \ref{table_1} presents a log of the observations and the data from the instruments that are used in this work.
{\bfseries{{\em Swift}-UVOT}}: The observations started $T_{0}$ + 84 s \citep{b11}. The afterglow is clearly detected. 

{\bfseries{TAROT La Silla}}: The observations at TAROT-La Silla \citep{boer_05} started at $T_{0}$ + 31.2 s and lasted for about 41 minutes, until the beginning
of sunrise \citep{b3}. The burst is not detected between 31 s and 68 s, with a limiting magnitude $R_{lim}$ = 16.6. After that time, the burst is clearly detected for the remainder of the observation. 

{\bfseries{Skynet PROMPT-CTIO}}: The observations with Skynet PROMPT-CTIO (two 14" telescopes), at Cerro Tololo, Chile \citep{rei05}, started at $T_{0}$ + 45 s and lasted for 27.25 m \citep{b4, b4b}. Forty-four exposures were taken in the V and I bands, ranging from 5s to 160s. 
The optical afterglow was clearly detected with a rising light curve at t = 2 min and peaks at I = 14.8. 
Skynet observed the afterglow again at $T_{0}$ + 23.0 h for 1.5 h, taking 64 exposures of 160s each in V and I bands. 
 
{\bfseries{GROND}}: GROND \citep{greiner_08} observations started at $T_{0} + 142 s$ \citep{b8}, and continued for 18 minutes. The afterglow was clearly detected. 

{\bfseries{KECK II telescope}}: Spectroscopic observations with the Keck II telescope were performed from $T_{0}$ + 1.78 h to $T_{0}$ + 2.15 h. Several lines were detected (Mg II doublet and Fe II), putting this burst at a redshift of z = 1.452 \citep{b9}.
 
\section{Data reduction}
\label{sec_red}
\subsection{Optical/IR data}
The TAROT data were reduced using the standard procedure already discussed in \citet{klotz_08}. We converted the observed signal from the clear filter to the R filter by calibrating the magnitude of the afterglow against nearby stars of similar color. 

\begin{figure*}
\centering 
\includegraphics[scale=0.8]{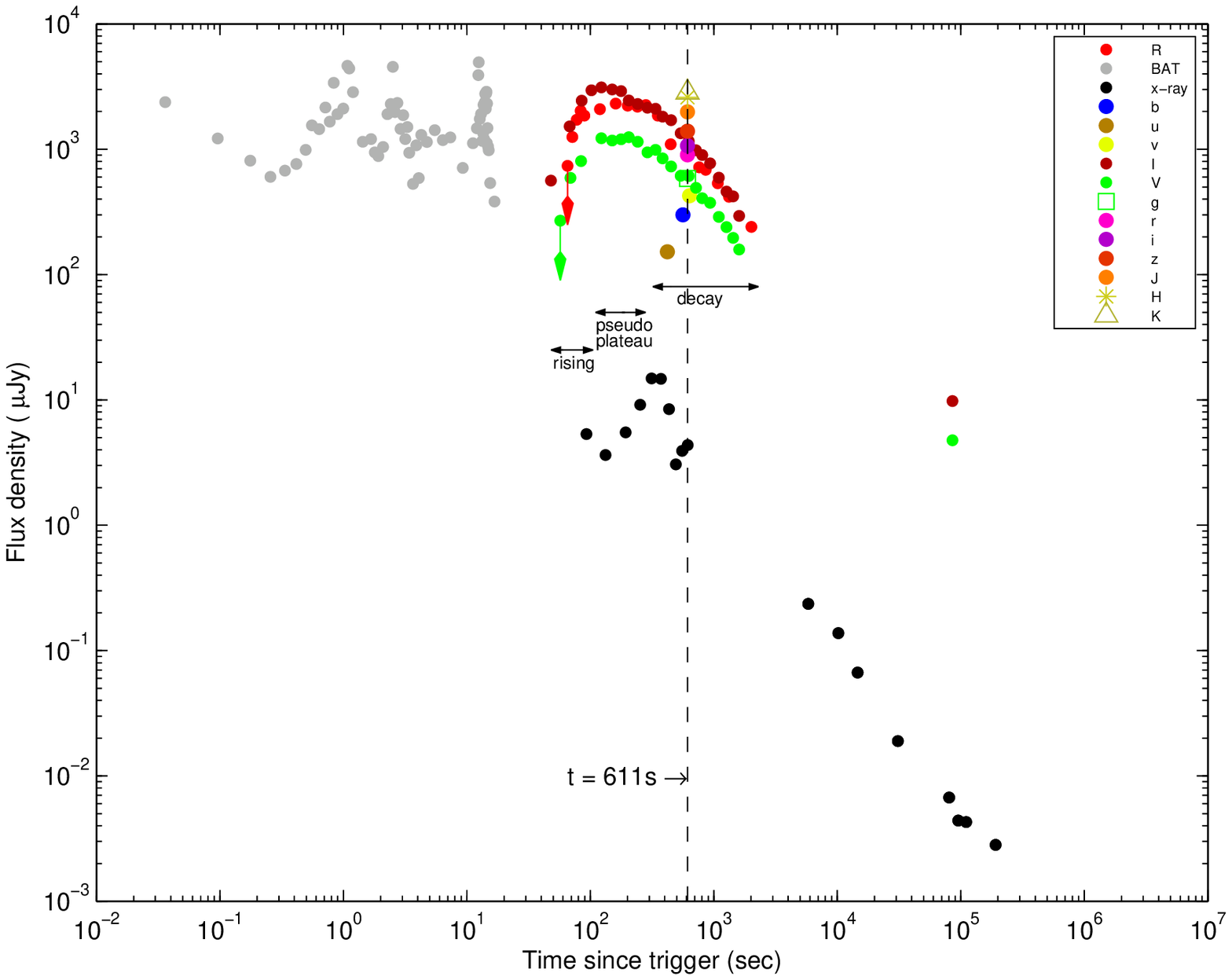}  
\caption {Flux density light curve of GRB 141221A. The vertical dashed line represents the epoch when the SED was extracted (see text for details).\label{fig2}}
\end{figure*}

Subsets of the Skynet images were stacked to maximize the signal-to-noise ratio. Calibration of these images was performed using three stars in the field from the AAVSO APASS DR7 catalog. The BVg'r'i' magnitudes from APASS were converted to BVRI Vega magnitudes using transformations provided by AAVSO (A. Hendon, private communication). Standard bias, dark, and flat corrections were applied to all images. Consecutive images were grouped and stacked in a way which maximizes the SNR of the afterglow while minimizing the loss of temporal resolution. The afterglow and a single primary calibration star were photometered in each stacked image and the resulting calibration offset was recorded.   A master calibration stack was then generated for each filter by combining all available images. For each master calibration stack, the primary calibration star was photometered as well as the two secondary calibration stars. By comparing the offset obtained from the secondary calibration stars to that obtained from the primary calibration star, a calibration correction is calculated and applied to all afterglow photometry.
The remaining data have been gathered from the literature and are compiled in Table \ref{table_1}. Fig. \ref{fig2} displays the resulting light curves.

All magnitudes were then converted into the AB system, if required. The correction for the Galactic extinction was applied at the same time, using a value of E(B-V) = 0.024 \citep{schlafly_11}. The reddening due to the host galaxy is left as free parameter in fits to be discussed below. This leads to the corrections listed in Table \ref{table_2}. We then computed from the corrected magnitudes the flux density, using a zero point value of 23.926. The final flux density light curves are presented in Fig. \ref{fig2}. 

\begin{table}
 \centering
  \caption{Corrections to magnitudes due to Galactic extinction. \label{table_2}}
  \begin{tabular}{cc}
  \hline \hline
Filter & Correction\\
\hline
u & 0.117\\
b & 0.092\\
V & 0.075\\
v' & 0.104\\
g' & 0.091\\
R & 0.074\\
r' & 0.063\\
I & 0.041\\
i' & 0.047\\
z' & 0.035\\
J & 0.019\\
H & 0.012\\
K & 0.080\\
\hline
\end{tabular}
\end{table}

\begin{table*}
\centering
 \begin{minipage}{140mm}
  \caption{Optical data converted into the AB System and corrected for Galactic extinction.\label{table_1}}

 \begin{tabular}{@{} r c c c c  | c c c c c @{}}
  \hline \hline
 
   Mid time & Filter & Magnitude & Telescope & Reference\footnote{\label{foot}References for the data: (1)~this work, (2)~\cite{b8}, (3)~\cite{b11} } & $\qquad$ Mid time & Filter & Magnitude & Telescope & Reference$^{\ref{foot}}$ \\
 
      (sec)    &              & AB System &                     &                     & $\qquad$ (sec)        &            & AB System &                     &  \\
   
 \hline
 65.46 & R & $<16.76$     &TAROT & (1) & $\qquad$ 57.00 & V & $<16.72$                 & Skynet & (1) \\ 
71.46 & R & 16.18$\pm0.2$ & TAROT & (1) & $\qquad$ 69.00 & V & $17.00^{+0.95}_{-0.54}$ & Skynet & (1) \\
77.46 & R & 15.84$\pm0.2$ &TAROT & (1) & $ \qquad$  84.00 & V & $16.66^{+0.24}_{-0.19}$ & Skynet & (1) \\ 
83.46 & R & 15.66$\pm0.2$ & TAROT & (1) & $ \qquad$ 101.00 & V & $15.93^{+0.12}_{-0.10}$ & Skynet &(1)\footnote{This point exhibits an instrumental bias and has not been included in the analysis} \\
89.46 & R & 15.76$\pm0.30$ &TAROT & (1) & $ \qquad$ 123.00 & V & $16.21^{+0.08}_{-0.08}$ & Skynet & (1) \\
119.60 & R & 15.63$\pm0.03$ & TAROT & (1) & $\qquad$ 150.00 & V & $16.25^{+0.08}_{-0.08}$ & Skynet & (1) \\
160.10 & R & 15.52$\pm0.03$ & TAROT & (1) & $ \qquad$ 177.00 & V & $16.23^{+0.08}_{-0.07}$ & Skynet & (1) \\
200.70 & R & 15.56$\pm0.03$ & TAROT & (1) & $\qquad$ 205.00 & V & $16.19^{+0.08}_{-0.07}$ & Skynet & (1) \\
241.00 & R & 15.58$\pm0.03$ & TAROT & (1) & $\qquad$ 242.00 & V & $16.28^{+0.05}_{-0.05}$ & Skynet & (1) \\
281.30 & R & 15.55$\pm0.03$ & TAROT & (1) & $ \qquad$ 290.00 & V & $16.49^{+0.07}_{-0.06}$ & Skynet & (1) \\
351.70 & R & 15.76$\pm0.09$ & TAROT & (1) & $ \qquad$ 337.00 & V & $16.44^{+0.06}_{-0.05}$ & Skynet & (1) \\ 
446.30 & R & 16.33$\pm0.02$ & TAROT & (1) & $ \qquad$ 384.00 & V & $16.61^{+0.07}_{-0.06}$ & Skynet & (1) \\ 
611.00 & r' & 16.54$\pm0.1$ & GROND & (2) & $ \qquad$  451.00 & V & $16.77^{+0.04}_{-0.04}$ & Skynet & (1) \\ 
760.60 & R & 16.79$\pm0.02$ & TAROT & (1) & $ \qquad$ 539.00 & V & $16.96^{+0.05}_{-0.05}$ & Skynet & (1)\\ 
861.20 & R & 16.84$\pm0.08$ & TAROT & (1) & $ \qquad$  627.00 & V & $16.96^{+0.05}_{-0.05}$ & Skynet & (1) \\ 
1074.40 & R  & 17.11$\pm0.08$ & TAROT & (1) & $ \qquad$ 636.00 & V & $17.35^{+0.26}_{-0.26}$ & UVOT & (3) \\
1327.50 & R & 17.38$\pm0.08$ & TAROT & (1) & $ \qquad$ 716.00 & V & $17.19^{+0.06}_{-0.06}$ & Skynet & (1) \\ 
2011.00 & R & 17.98$\pm0.08$ & TAROT & (1) & $ \qquad$ 804.00 & V & $17.40^{+0.07}_{-0.07}$ & Skynet & (1) \\
48.00 & I & $17.05^{+0.65}_{-0.42}$ & Skynet & (1) & $\qquad $ 931.00 & V & $17.49^{+0.06}_{-0.05}$ & Skynet & (1) \\ 
68.00 & I & $15.97^{+0.12}_{-0.11}$ & Skynet & (1) & $\qquad $ 1098.00 & V & $17.77^{+0.07}_{-0.07}$ & Skynet & (1) \\ 
85.00 & I & $15.46^{+0.07}_{-0.06}$ & Skynet & (1) & $\qquad $ 1266.00 & V & $17.98^{+0.10}_{-0.09}$ & Skynet & (1) \\ 
102.00 & I & $15.25^{+0.06}_{-0.06}$ & Skynet & (1) & $\qquad $ 1433.00 & V & $18.19^{+0.13}_{-0.12}$ & Skynet & (1) \\ 
123.00 & I & $15.19^{+0.04}_{-0.03}$ & Skynet & (1) & $\qquad$ 1600.00 & V & $18.43^{+0.17}_{-0.15}$ & Skynet & (1) \\ 
150.00 & I & $15.23^{+0.03}_{-0.03}$ & Skynet & (1) & $\qquad$  85523.00 & V & $22.23^{+3.83}_{-1.10}$ & Skynet & (1) \\ 
177.00 & I & $15.26^{+0.03}_{-0.03}$ & Skynet & (1) & $\qquad$ 611.00 & g' & 17.01$\pm0.1$ & GROND & (2) \\ 
205.00 & I & $15.45^{+0.04}_{-0.04}$ & Skynet & (1) & $\qquad$ 611.00 & z' & 16.07$\pm0.1$ & GROND & (2) \\ 
242.00 & I & $15.52^{+0.03}_{-0.03}$ & Skynet & (1) & $\qquad$  611.00 & J & 15.68$\pm0.1$ & GROND & (2) \\ 
290.00 & I & $15.60^{+0.03}_{-0.03}$ & Skynet & (1) & $\qquad$ 611.00 & H & 15.39$\pm0.1$ & GROND & (2) \\
337.00 & I & $15.62^{+0.03}_{-0.03}$ & Skynet & (1) & $\qquad$ 611.00 & K & 15.29$\pm0.1$ & GROND & (2) \\ 
384.00 & I & $15.77^{+0.04}_{-0.04}$ & Skynet & (1) & $\qquad$  561.50 & b & 17.73$\pm0.21$ & UVOT & (3) \\ 
451.00 & I & $15.85^{+0.02}_{-0.02}$ & Skynet & (1) & $\qquad $ 421.00 & u & 18.47$\pm0.07$ & UVOT & (3) \\ 
539.00 & I & $16.11^{+0.03}_{-0.03}$ & Skynet & (1) & $\qquad$ & & & & \\  
611.00 & i' & 16.35$\pm0.1$& GROND & (2) & $\qquad$ & & & & \\ 
627.00 & I & $16.26^{+0.04}_{-0.03}$ & Skynet & (1) & $\qquad$ & & & &\\ 
716.00 & I & $16.45^{+0.04}_{-0.04}$ & Skynet & (1) & $\qquad$ & & & &\\ 
804.00 & I & $16.54^{+0.04}_{-0.04}$ & Skynet & (1) & $\qquad$ & & & &\\ 
931.00 & I & $16.71^{+0.03}_{-0.03}$ & Skynet & (1) & $\qquad$ & & & & \\ 
1098.00 & I & $16.99^{+0.04}_{-0.04}$ & Skynet & (1) & $\qquad$ & & & & \\ 
1265.00 & I & $17.27^{+0.06}_{-0.06}$ & Skynet & (1) & $\qquad$ & & & &\\ 
1433.00 & I & $17.36^{+0.07}_{-0.06}$ & Skynet & (1) & $\qquad$ & & & &\\ 
1600.00 & I & $17.75^{+0.10}_{-0.09}$ & Skynet & (1) & $\qquad$ & & & &\\ 
85609.00 & I & $21.45^{+0.67}_{-0.43}$ & Skynet & (1) & $\qquad$ & & & &\\ 
\hline
\end{tabular}
\end{minipage}
\end{table*}

\subsection{Fermi data}
GBM data for GRB 141221A were downloaded from the NASA/GSFC Fermi GBM Archive. The extraction of GBM data was done by using only the NaI detectors with the brightest signal in the 8keV -- 1MeV band. In the case of GRB 141221A, these detectors were NaI 01 and NaI 02. We used the task {\tt \small RMFIT(v432)} for data reduction, using the event files (TTE) of each good detector.

As the high-energy light curve of this event consists of two pulses, we performed all analysis both on each pulse separately and on the full time interval to check for spectral variations. For that purpose, we used the 8.0 -- 900.0 keV energy band.

\subsection{X-Ray data}
The data for GRB 141221A were downloaded from the NASA/GSFC {\em Swift} Data Center
and were processed using {\tt \small{HEASoft(v6.16)}} and the XRTDAS software version 0.13.1, with the latest calibration files available in June 2015.  
We used the task {\tt \small xrtpipeline} to create the clean event file and to apply the latest calibration. 
We then performed a screening for bad pixels and piled-up data, using the methods and corrections indicated in \citep{romano_06} and \citep{vaughan_05}. We found that the flare observed in PC mode is piled-up during the interval $T_0$ + 138.2 s -- $T_0$ + 619.7 s. Last, we restricted the analysis to events with energy between 0.3 and 10.0 keV. This led to a net exposure of 50.53 s in the Window Timing mode (hereafter WT) and 26251.72 s in the Photon Counting mode (PC).

\section{Data analysis}
\label{sec_ana}
\subsection{Prompt data}
As already indicated, the prompt light curve has a duration (T$_{90}$) of about 23.8 seconds in Fermi-GBM and about 37s in Swift-BAT, and consists of two pulses. For the spectral analysis of prompt emission we used the Fermi/GBM instead of {\em Swift}/BAT data because of the much larger energy band of the former instrument. We used Xspec version 12 \citep{arn96} to fit the spectrum with a Band model \citep{band_93}. We first fit each pulse separately (named Intervals 1 and 2, respectively), and then fit the complete spectrum. We also took an average of the two pulse results. All the results are displayed in Table \ref{table3}, together with a reminder of the GCN result \citep{b12}. The low signal prevented us from fitting all the Band parameters separately, and in all cases we had to fix the $\beta$ parameter to a value of -2.3.

\begin{figure}
\centering 
\includegraphics[width=8cm]{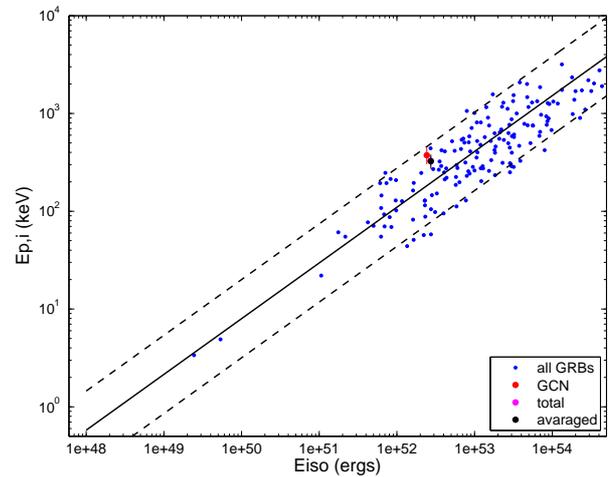}  
\caption {Our GRB compared to the whole sample of GRBs until June 2013. The solid line is $E_{p,i} = 110*E_{iso}^{0.57}$, while the dashed line is the 2 sigma standard deviation \citep{amati_09}.\label{fig3}}
\end{figure}

\begin{table*}
\centering
\caption{\label{table3} Results of the prompt spectral fitting. Non-constrained parameters are fixed to the values indicated in square brackets}.
\begin{tabular}{ccccccccc}
\hline \hline
Interval & time & exposure time     & $\alpha$           & $\beta$ & C-Stat & d.o.f & $E_{p,i}$      & $E_{iso}$ \\ 
         & (sec) & (sec)            &                    &         &        &       & (keV)        &($10^{52}$ ergs) \\
\hline
1        & -1.024 -- 8.704 & 9.728 & [-1.00]            & [-2.30] & 431.57 & 241 & 353 $\pm$ {42} & 1.88 $\pm$ {0.11} \\
2        & 8.704 -- 17.408 & 8.704 & -0.82 $\pm$ {0.38} & [-2.30] & 523.84 & 253 & 247 $\pm$ {77} & 0.83 $\pm$ {0.09} \\
2        & 8.704 -- 17.408 & 8.704 & [-1.00]            & [-2.30] & 523.98 & 254 & 297 $\pm$ {61} & 0.88 $\pm$ {0.10}  \\
Total    & -1.024 -- 17.408 & 18.432 & -1.24 $\pm$ {0.11} & [-2.30] & 558.20 & 240 & 531 $\pm$ {164} & 3.13 $\pm$ {0.25} \\
Total & -1.024 -- 17.408 & 18.432 & [-1.00]            & [-2.30] & 558.97 & 241 & 328 $\pm$ {35} & 2.71 $\pm$ {0.14} \\
averaged & -1.024 -- 17.408 & 18.432 & [-1.00]            & [-2.30] & 477.77 & 247.5 & 325 $\pm$ {52} & 2.74 $\pm$ {0.21} \\
\hline
GCN      & -1.024 -- 17.408 & 18.432 & -1.07 $\pm$ {0.13} & ---     & ---    & ---   & 374 $\pm$ {70} & 2.43 $\pm$ {0.29} \\
\hline
\end{tabular}
\end{table*}
Knowing $E_{peak}$ and the distance of this burst, we have calculated $E_{p,i} = 374 \pm{70}$ keV and $E_{iso} = 2.4 \times 10^{52}$ erg. We note that these values follow the $E_{p,i} - E_{iso}$ relation \citep{amati_09, amati_02}, as can be seen in Fig. \ref{fig3}.

\subsection{Temporal decay}
\subsubsection{X-ray}
The X-ray temporal analysis was already done for the extraction of the SED. The light curve presents a prominent flare, peaking at about 340 seconds. The remainder of the afterglow light curve is well fit by a simple power law, as can be seen in Table \ref{table6} and Fig. \ref{fig2}.

\begin{table*}
 \centering
  \caption{Best fit temporal decay indices for the I, R, V and X-ray bands. Numbers in parentheses are not constrained by the fit. See text for details.\label{table6}}
  \begin{tabular}{cccccccccc}
  \hline \hline
time        & filter & model              & $\alpha_{1}$ & $\alpha_{2}$ & $\alpha_{3}$ & $t_{break}$    & $t_{break,2}$ & $\chi^2_\nu$ & d.o.f \\
(sec)       &        &                    &                &                &               & (sec)          & (sec)         &              & \\
\hline
 48 - 337 & I      & broken power law & $-1.6 \pm 0.9$ & $0.5 \pm 0.2$ & ---           & $110 \pm 13$ & ---            & 1.67         & 6  \\
337 - 85609 & I      & broken power law & $1.0 \pm 0.2$ & $1.6 \pm 0.4$ & ---           & $918 \pm 160$ & ---            & 1.53         & 9 \\
281 - 2011 & R      & 2 broken power law & $1.6 \pm 0.4$ & $0.4 \pm 3.2$ & $1.3 \pm 0.9$ & $540 \pm 514$ & $906 \pm 696$ & 1.36         & 1 \\
 69 - 205 & V      & broken power law & $(-1.6)$       & $0.1 \pm 1.2$ & ---           & (109)          & ---            & 0.10         & 1  \\
205 - 85523 & V      & broken power law & $0.7 \pm 0.1$ & $1.3 \pm 0.3$ & ---           & $641 \pm 125$ & ---            & 0.70         & 12 \\
5800 - 191504 & X-ray& power law          & $1.4 \pm 0.2$ & ---            & ---           & ---            & ---            & 1.31         & 6 \\
\hline
\end{tabular}
\end{table*}

\subsubsection{Optical}

\begin{table}
 \centering
  \caption{Simple power law decay fit of the I, R, V bands. See text for details.\label{table7}}
  \begin{tabular}{cccc}
  \hline \hline
Filter & $\alpha$ & $\chi^2_\nu$ & d.o.f \\
\hline
I        & $1.12 \pm {0.10}$ & 3.83 & 11 \\
V        & $0.91 \pm {0.10}$ & 2.57 & 14 \\
R        & $1.04 \pm {0.22}$ & 10.6 & 6 \\
\hline
\end{tabular}
\end{table}

The optical light curves are more complex than the X-ray one. They present a rise, a pseudo-plateau, and a decay. We split the study in two parts, namely the rising and the decaying parts.

For the rise, we used a broken power law model. This gives us the end time of the fast rise and the start of the pseudo-plateau phase. In a few cases, the lack of data prevented an accurate measure, and we indicate these as numbers in parentheses in Table \ref{table6}. This is the case for the R band, which we attribute to an instrumental bias (see below).

For the decay, we first tried a simple power law model. As one can see in Table \ref{table7}, this model is strongly rejected in all bands. We then inserted a break in the power laws, obtaining good fits in the V and I bands (see Table \ref{table6}). However, this model, surprisingly, still does not fit the R band. In that band, we need a double broken power law in order to obtain a correct fit. At that point, the degrees of freedom are too low to ensure a correct measurement of the errors.

\begin{figure*}
\includegraphics[width=8cm]{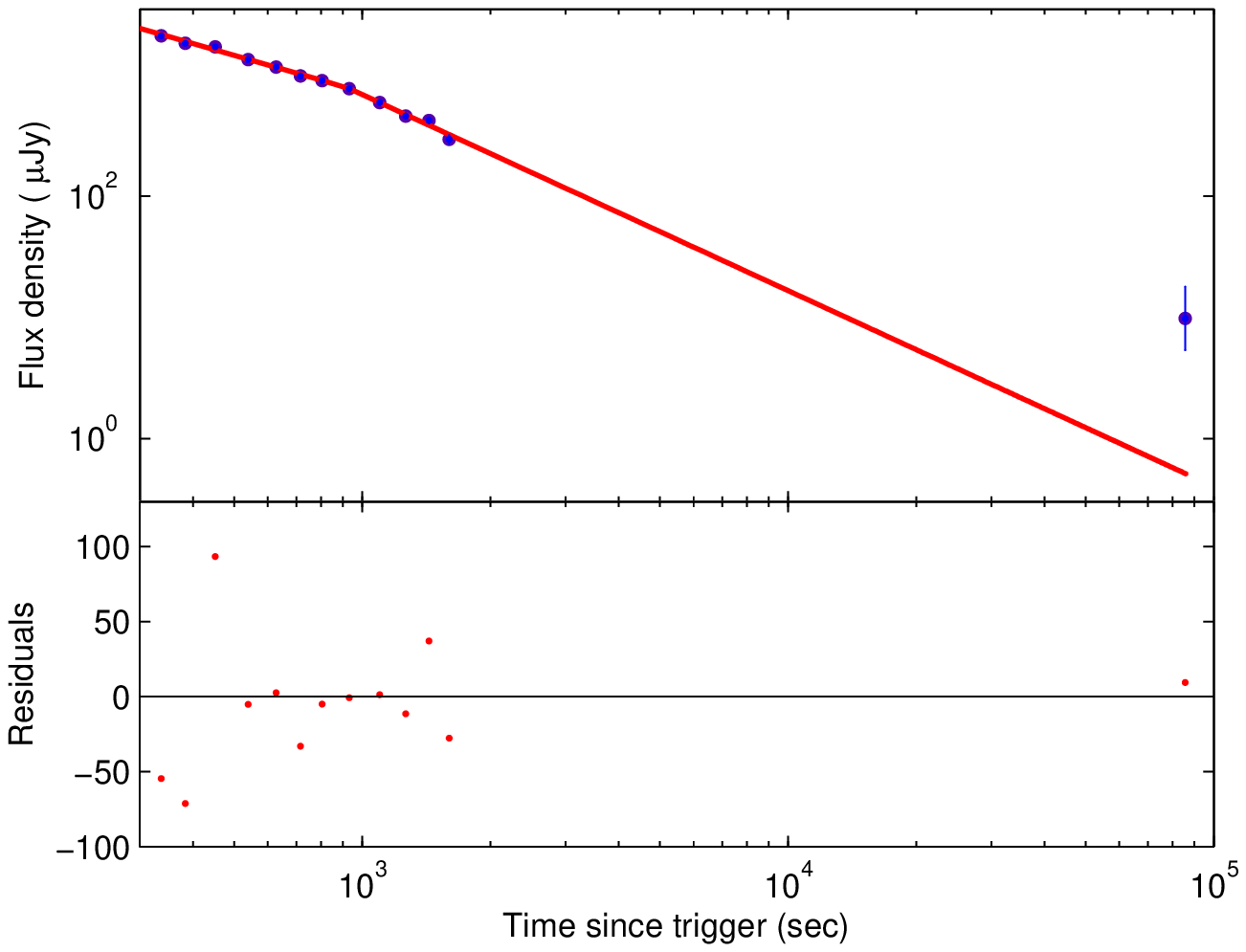}
\includegraphics[width=8cm]{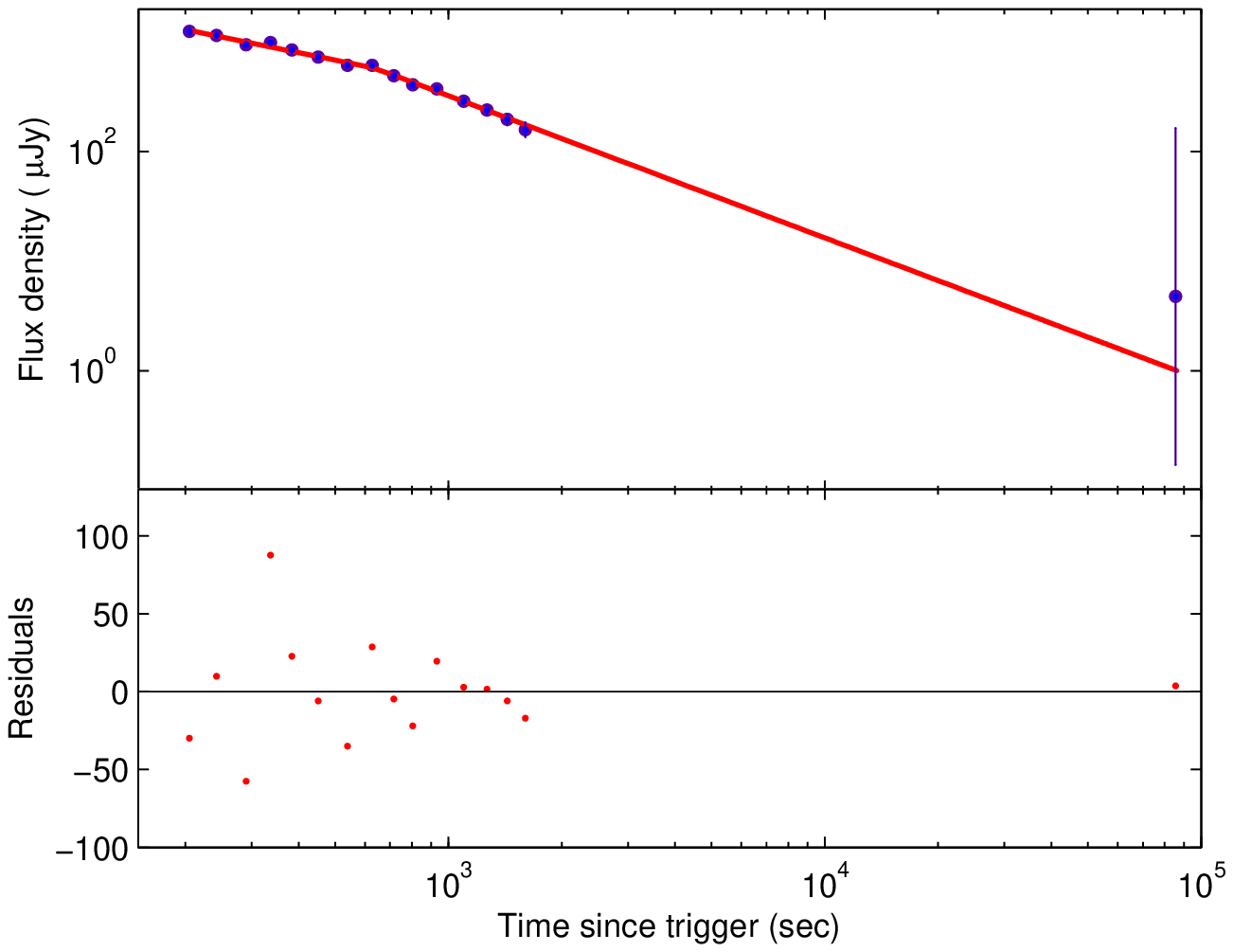}
\caption {The best fit in the I (left) and V (right) bands with a power-law decay, starting from the end of the pseudo-plateau. The lower parts of each figure show the residuals of the fits.\label{fig_fit}}
\end{figure*}

This double broken power law model mimics the standard Swift X-ray light curve (i.e. a steep-flat-steep shape), but is not seen in the other bands. We explain this feature by the fact that these R band measurements come from the TAROT telescope, which was unfiltered to maximize its sensitivity. We have normalized the magnitudes to the Cousin R band assuming a template afterglow spectrum that does not contain any break. The TAROT CCD camera is sensitive from the I to the V bands (the B sensitivity is very low). A spectral break that appears partly in the observation window will not be accounted for. This can introduce an error in the reduced R magnitude that will depend on the position of the break. If the break is in the blue part of the spectrum, then the R magnitude will be underestimated, and vice-versa for the opposite case. The crossing of a spectral break would then translate into a steep-flat-steep shape in the light curve during the whole time of the crossing. This is not observed for the other bands (I and V) as standard filters have been used. The fits in the V and I bands (decay) are presented in Fig. \ref{fig_fit}.

\subsection{Afterglow spectrum}
\begin{figure}
\includegraphics[width=8cm]{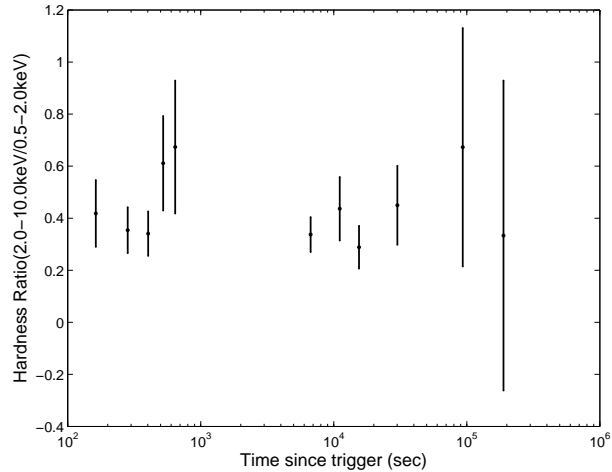}
\caption{\label{fig4}Hardness ratio of the X-ray observation. We used the hard and soft bands of 2.0-10.0 keV and 0.5-2.0 keV respectively, only in PC mode.}
\end{figure}

We started by analyzing the XRT spectrum alone, independently of the optical data. This is because at high energy (above 2 keV), the spectrum is not influenced by the surrounding medium and the column density, and thus the X-ray spectrum allows us to derive the intrinsic power law index. We extracted three spectra, one in WT mode and two in PC mode (during the flare, and after the flare), and fit these with a power-law model absorbed twice (one let free to vary at the distance of the burst, the second fixed to the galactic value in the direction of the burst, $N_{H}^{gal} = 2.27 \times10^{20} cm^{-2}$). The data are consistent with no spectral variation, though we note that the error bars are large due to the low flux of the afterglow. The results of these fits are presented in Table \ref{tab:xrt_results}. The lack of spectral variation is clearly confirmed by an analysis of the hardness ratio (using the hard and soft bands of 2.0-10.0 keV and 0.5-2.0 keV respectively) presented in Fig. \ref{fig4}. While we see at the end of the flare a possible hardening of the spectrum, the error bars are still consistent with no spectral variation at the $3\sigma$ level.

\begin{table}
 \centering
  \caption{X-ray spectral analysis, independent of the optical measurements. See text for details.\label{tab:xrt_results}}
  \begin{tabular}{cccccc}
  \hline \hline
Interval     & mode & $N_{H}^{host}$         & $\beta_{x}$         & $\chi^2_\nu$ & d.o.f.\\
(sec)        &      & ($10^{22}$ cm$^{-2}$) &                     &              &       \\
\hline
60 - 90      & WT & $0.27^{+2.3}_{-0.27}$ & $0.7^{+0.7}_{-0.5}$ & 1.02         & 6     \\
100 - 1000 & PC & $0.9^{+0.5}_{-0.4}$ & $1.0^{+0.4}_{-0.4}$ & 0.96         & 15    \\
3000 - 11000 & PC & $0.5^{+0.6}_{-0.4}$ & $1.0^{+0.4}_{-0.4}$ & 0.89         & 7     \\
\hline
\end{tabular}
\end{table}

Once we had the information on the power law spectral index at high energy, we built the Spectral Energy Distribution (SED), this time using all data available. We extracted the SED where the data were the most numerous, at about 611 s post burst (see Fig.~\ref{fig2}). This corresponds to the end of the flare in X-ray and the decay phase of the optical band. In X-ray, we used the data taken between 350 and 619.7 seconds, and normalized them to the underlying afterglow flux. This last point is important: as there is no hint of flare in the optical light curve, it should not be linked to the X-ray flare. The non-variability of the hardness ratio makes us confident that this renormalization is enough to correct for the presence of the flare. All data (including the optical data) were then imported into XSPEC for the spectral fitting.

\begin{figure*}
\centering 
\includegraphics[width=8cm]{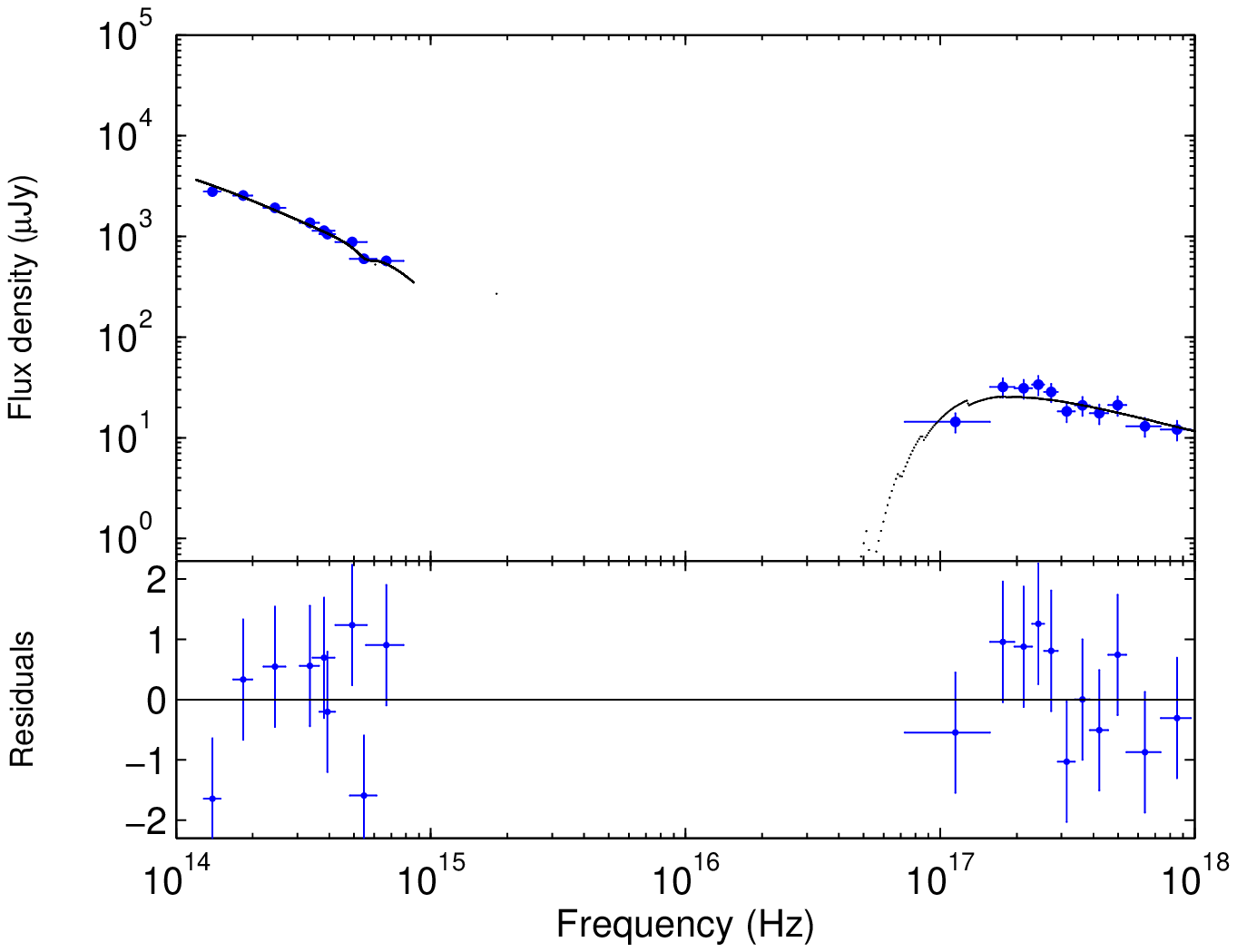}  
\includegraphics[width=8cm]{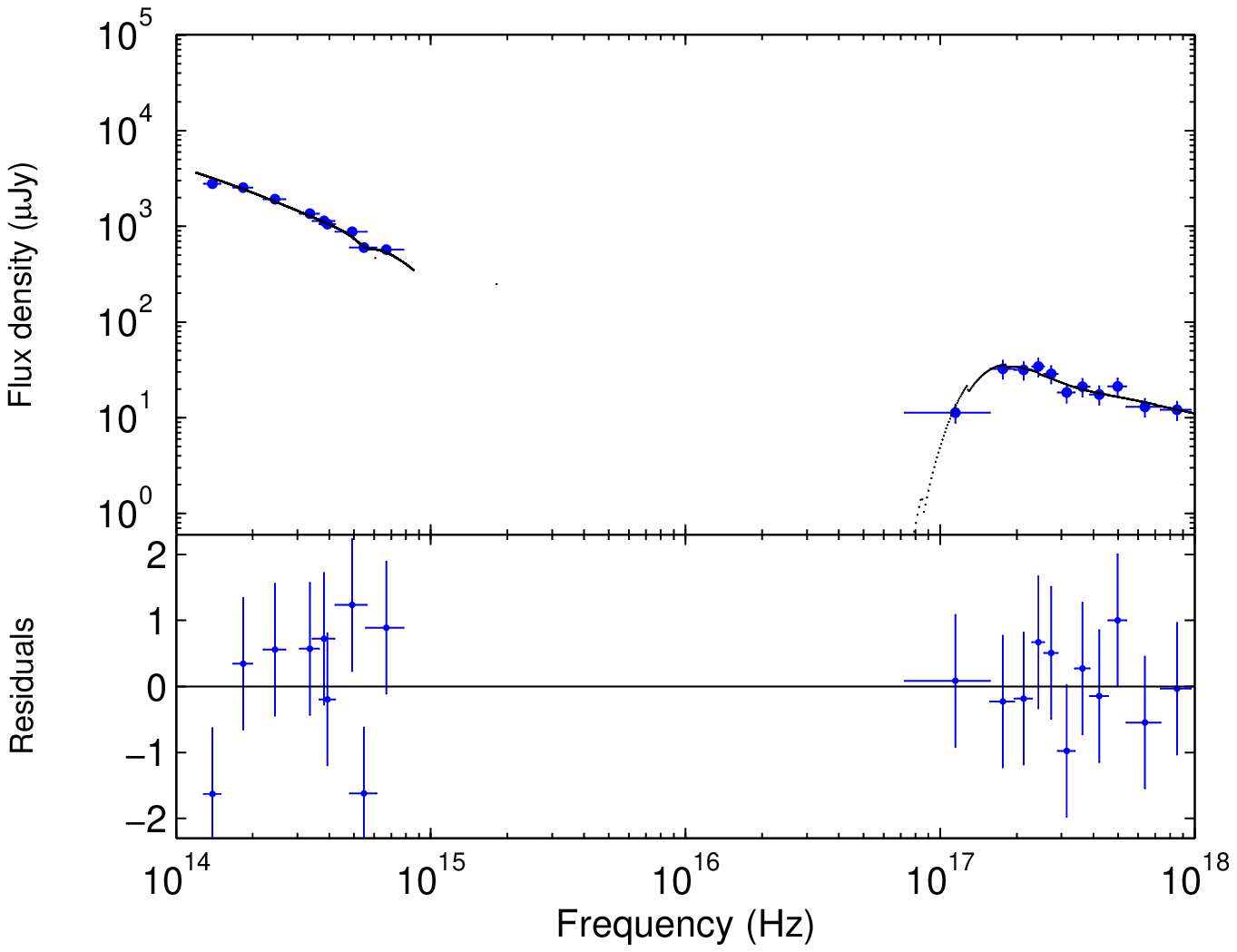}  
\caption {The SED of GRB 141221A, fit with various models. On the left, a simple power law model. On the right, a simple power law plus a thermal component. In both cases, we use the LMC extinction law to fit the optical data. The bottom panels show the residuals for each model.\label{fig5}}
\end{figure*}

To model the SED, we consider single power law, double power law and thermal components (see Table \ref{table5}). In all cases, we added foreground absorption by our own Galaxy \citep[this absorption was fixed to the measured values of][the optical extinction being corrected before the insertion into XSPEC]{kal05}, and by the distant host galaxy. We consider the three standard extinction laws, i.e., the Milky Way (MW), the Large Magellanic Cloud (LMC) and the Small Magellanic Cloud (SMC) ones. In all cases, the high energy power law index was allowed to vary freely only within the measured X-ray confidence interval. We first considered a simple power law extincted model. Even if the fit quality seems good (see Table \ref{table5}), an analysis of the residuals shows that this model does not fit the data correctly: it exhibits a lack of emission in the soft X-ray part of the SED (see Fig. \ref{fig5}). We then inserted a thermal component into the model, and redid the analysis. This time, both the quality indicator of the fit and the residuals are in agreement with a good solution. We also tested the hypothesis of a cooling break, i.e., a broken power law with the two spectral indices linked together by a difference of $\beta_{X} = \beta_{o} + 0.5$, which also provides an acceptable fit.

\begin{table*}
\centering
\caption{\label{table5} Results of the spectral analysis of the SED. $\beta_{o}$ is the power law index in case of a single power law. In case of a broken power law, this is the spectral index of the low energy segment, the high energy segment being linked to it by the relation $\beta_{x}$ = $\beta_{o}$ + 0.5. See text for details.}
\begin{tabular}{c c c c c c c c c c c c}
  \hline \hline
Model     & Extinction & $N_{H, host}$           & $R_{V}$ & $E(B-V)$            & $\beta_{o}$            & Temperature or         & $\chi^{2}_\nu$ & d.o.f \\
          &       &                              &         &                     &                        & break energy           &                &  \\
          & law & ($\times 10^{22}$ cm$^{-2}$) & (mag) & (mag)               &                        & (keV)                  &                &       \\
\hline
pow       & MW    & $0.4 \pm 0.3$                & 3.08 & $0.12 \pm 0.02$        & $0.63^{+0.03}_{-0.02}$ & ---                    & 0.7674 & 16 \\
          & LMC & $0.4 \pm 0.3$                & 3.16 & $0.12 \pm 0.02$        & $0.63 \pm 0.02$        & ---                    & 0.9855 & 16 \\
       & SMC & $0.4 \pm 0.3$                & 2.93 & $0.12 \pm 0.02$        & $0.63^{+0.03}_{-0.02}$ & ---                    & 1.3111 & 16 \\
         
\hline
pow+bbody & MW    & $1.3^{+1.9}_{-1.0}$          & 3.08 & $0.11 \pm 0.02$        & $0.63 \pm 0.03$        & $0.14^{+0.17}_{-0.04}$ & 0.607 & 14 \\
          & LMC & $1.3^{+1.9}_{-1.0}$          & 3.16 & $0.12 \pm 0.02$        & $0.64 \pm 0.03$        & $0.13^{+0.16}_{-0.05}$ & 0.862 & 14 \\
     & SMC & $1.3 \pm 0.9$                & 2.93 & $0.12 \pm 0.02$        & $0.63 \pm 0.03$        & $0.14^{+0.17}_{-0.05}$ & 1.229 & 14 \\
                                                  
\hline
cooling & MW    & $0.8^{+0.5}_{-0.4}$          & 3.08 & $0.14^{+0.05}_{-0.04}$ & $0.5 \pm 0.2$          & $<0.17$                 & 0.645 & 15 \\
break     & LMC & $0.6^{+0.5}_{-0.3}$          & 3.16 & $0.18^{+0.02}_{-0.06}$ & $0.3^{+0.3}_{-0.1}$    & $0.012^{+0.8}_{-0.01}$ & 0.7 & 15 \\
     & SMC & $0.7 \pm 0.3 $               & 2.93 & $0.17^{+0.03}_{-0.05}$ & $0.37^{+0.03}_{-0.09}$ & $0.03^{+1.6}_{-0.028}$ & 1.15 & 15 \\
\hline
\end{tabular}
\end{table*}

As can be seen, the addition of the optical data strongly constrains the spectral index of the power law to a very low value. On the other hand, from this fit we cannot discriminate between a Galactic or a Large Magellanic Cloud law of extinction. We present the best fit SED (assuming a power law model with an additional thermal component) in Fig. \ref{fig5}, using the LMC law, which is more common for GRBs compared to the MW law \citep{str04}.

\section{Discussion}
\label{sec_discu}
\subsection{The thermal component}
We first consider the possibility that the thermal component seen in the SED is real. This would not be the first time such a component has been observed in the Swift era \citep{rha12, spa12}. It has been explained either as the shock breakout of the supernova onto the surface of the progenitor or the emission of a hot cocoon protecting the jet during its travel into the progenitor \citep{but07}. We note incidentally that this last explanation was also proposed to describe the early emission of ultra-long GRBs \citep{gen13, pir13}, even if, as in this case, the burst does not belong to that class of events. As no supernova has been reported for GRB 141221A, we do favor the hypothesis of the hot cocoon.

If this component is really present, then the SED indicates that the optical and X-ray emissions are linked together, and are thus due to the same emission mechanism. Indeed, at late times, all the temporal decay indices are compatible, within errors. However, the SED is extracted before the final break of the I band, and thus this should also apply to earlier measurements. We do not see any evidence of a break in the X-ray light curve: this can be explained by the presence of the flare, which masks out the actual evolution of the afterglow. Moreover, the break times of the I and V magnitudes are compatible, within errors.

This light curve break is then achromatic, which is consistent with a jet break \citep{rho97, rho99}. We obtain a value of $p= 1.28 \pm 0.06$ which is extremely low. In addition, the jet break time is also extreme \citep[about 750 s, while a common pre-Swift value is on the order of days][]{gen07}. This would lead to a jet opening angle of 1.3 degrees \citep[assuming the standard law of][]{sar99}, and could explain why in most cases no jet break is observed for Swift bursts: the break is looked for around a few hours (or days) after the trigger, and not at that earlier time.

In addition to this surprising value of the jet opening angle (that would put strong constraints on the SFR of massive stars in the Universe), the only argument against this hypothesis is the R band behavior, that does not follow the V and I bands. In the previous section, we have explained this behavior by the fact that the break time was not identical in all bands. If we suppose a constant break time, we cannot explain the R band behavior.

\subsection{The rising and early decay of the afterglow}
Let us now assume that the thermal component is not real, and instead use a broken power model for the SED. At a late time, all the temporal decay indices are compatible, within errors. Now, the SED tells us that the X-ray and optical emissions are not linked to the same emission mechanism at the time of the SED (611 s). We can then assume that the various breaks we see are due to the passing through of a specific frequency into the observation bands, and that at a late time (> about 1000 s), the crossing of this frequency has ended and all the emission is due to the same emission mechanism.

The temporal break times in the I and V bands indicate that this specific frequency is decreasing with time, and, as already explained, the R band behavior is also compatible with that hypothesis. This leads us to exclude the passing through of the cooling frequency in a wind medium, as this frequency increases with time in such a case \citep{che99,pan00}. If we still assume the wind medium, the only remaining option is the injection frequency, $\nu_m$. However, the spectral index before the crossing ($0.3^{+0.3}_{-0.1}$) would lead to a value of $p$ lower than 1, which is not physical. We thus can conclude that these breaks cannot be explained in the case of the wind medium.

The situation is different in case of the ISM. There, we can logically assume that the last two breaks are linked to the injection and cooling breaks, respectively. The injection and cooling frequencies vary as $t^{-1.5}$ and $t^{-0.5}$ respectively. Taking into account the errors on the break times, all break measurements are compatible with this explanation. After the cooling break, the spectral and the temporal decay indices are all compatible with a value of $p \sim 2.5 \pm 0.3$. The early spectral index (before the cooling break, as measured in the optical) should be $\beta = 0.7 \pm 0.2$, compatible with the measurement ($0.3^{+0.3}_{-0.04}$).

In this scenario, the end of the "pseudo-plateau" phase is the injection break, i.e., the peak of the afterglow. Again, the variation of the break time between the V and I bands is consistent with this hypothesis. Then, however, the temporal decay indices of the "pseudo-plateau" should become negative. This does not agree with the model. We explain it by the contribution of a small reverse shock that masks the peak of the emission. We can then, assuming the surrounding medium density to be equal to one, and the efficiency of the fireball in radiating its energy to be 30 \%, compute the microphysical parameters of the fireball, using the work of \citet{pan00}. Doing so, we obtain the fireball total energy ($E = 8 \times 10^{52}$ erg), the magnetic parameter ($\epsilon_B = 5 \times 10^{-2}$), and the electron parameter ($\epsilon_e = 3 \times 10^{-3}$). These numbers are relatively normal \citep[see e.g.][]{gen08}, albeit $\epsilon_B$ is slightly higher than usually seen. We thus have a complete description of the afterglow of this burst. We note, however, the total absence of a stellar wind in that model.

\citet{che04} have pointed out the complex surrounding medium of a GRB. However, assuming that the progenitor for all long GRBs is a stellar object \citep{woo93}, we still should observe a small portion of the light curve where a wind environment should be present. Here, from about 200 seconds after the trigger to the end of the observations, the medium is compatible with an ISM only. It is a well-known fact that most of Swift bursts are compatible with an ISM, but a degeneracy prevents excluding the wind medium hypothesis \citep{che04}. Here, we have the proof that the wind medium is rejected from nearly the start of the afterglow, leaving only extreme constraints on the stellar physics in order to suppress the stellar wind from the progenitor. It is beyond the scope of this paper to introduce such a stellar model, however GRBs are known to have weak stellar winds \citep[e.g.][]{gen04, gen13}, and thus such a model would be very useful. We conclude this section by noting that the intrinsic values of $E_{B-V}$ and N$_H$ are low, and thus again are compatible with a low density around this GRB.

\subsection{Absorption and Extinction}
From our analysis, it turns out that we obtain a better solution using an LMC extinction law, because the observed GROND g-band is best fit by 2175 \AA~absorption 
feature present in LMC (and MW). We note that best-fit solutions with LMC or MW dust
have already been observed \citep[e.g.,][]{kan06,kru08,kan10}, even if other models may be more appropriate \citep{str04}.
However, given that these data were obtained from the preliminary photometry quoted in \citet{b8}, and that
they do not have appreciable influence on the fitted parameter values, we prefer to leave this argument for a future work when better data will be available.

All the spectral models we tried favor a slightly dusty environment with $E(B-V)\sim0.1-0.2$ (see Table \ref{table5}). These values are not unusual \citep{kan10, gre11, zaf11}, most of all at the distance of GRB 141221A \citep{kan06,cov13}. The observed $N_{H,host}$ is also in agreement with those found for other 
bright bursts, especially when compared with the best-fit optical extinction in the redshift interval $1<z<2$ \citep[e.g.][]{wat13, cov13}. Like many other bursts,
the metals-to-dust ratio ($N_{H,host}/A_V$) is in the range $1-3\times 10^{22}\, cm^{-2} \, mags^{-1}$ \citep{zaf11, kru11, cov13}.

We finally note that the extinction is not enough to set the optical to X-ray spectral index below the value $\beta_{\rm O-X}=0.5$ (see Table \ref{table5}), and thus we cannot consider GRB 141221A as a dark GRB \citep{jak04, ros12}.
 
\section{Conclusions}
We have analyzed the observations of GRB 141221A made in optical and high energy bands by various instruments, including TAROT and Skynet. In X-ray bands, the burst is very similar to all the previous ones observed, with a late flare. In optical bands, however, the light curve shows a rising part, a pseudo-plateau phase, and various temporal breaks. We explain these breaks as due to the passing through of several specific frequencies into the optical bands. We need a minimal contribution by a reverse shock to completely explain both the optical and X-ray light curves and spectra. 

An alternative hypothesis would be the presence of a thermal component, to explain the observed optical/X-ray SED. In this case, the last temporal break observed would be due to a jet effect. This, however, would lead to various properties being, while not formally forbidden by the model, extreme, and, in addition, would lead to the presence of a thermal emission in the soft X-ray band. All of these facts are unusual and difficult to explain.

Clearly, both solutions are challenging for GRB models. In the former case, all the data point toward an absence of stellar wind during the whole phenomenon, which is in contradiction with current models. In the latter case, the microphysics parameters obtained by the model are very unusual, and in some cases not really taken into account by the model. GRB 141221A should thus be added to the short list of very constraining bursts against which each new model should be tested.

\section*{Acknowledgments}
We thank the anonymous referee for her/his helpful comments that helped to improve this paper.
O.Bardho is supported by the Erasmus Mundus Joint Doctorate Program by Grant Number 2012-1710 from the EACEA of the European Commission. 
This research has made use of the
XRT Data Analysis Software (XRTDAS) developed under the responsibility
of the ASI Science Data Center (ASDC), Italy. BG acknowledges financial support of NASA through the NASA Award NNX13AD28A and the NASA Award NNX15AP95A. AR, and EP acknowledge support from PRIN-INAF 2012/13. This work is under the auspice of the FIGARONet collaborative network, supported by the Agence Nationale de la Recherche, program ANR-14-CE33.

\label{lastpage}

\begin{thebibliography}{}
\bibitem[\protect\citeauthoryear{Amati et al.}{2002}]{amati_02} Amati, L., Frontera, F., Tavani, M., 2002, A\&A 390, 81
\bibitem[\protect\citeauthoryear{Amati et al.}{2009}]{amati_09} Amati, L., Frontera, F., Guidorzi, C., 2009, A\&A 508, 173
\bibitem[\protect\citeauthoryear{Arnaud}{1996}]{arn96} Arnaud, K.A., 1996, Astronomical Data Analysis Software and Systems V, eds. Jacoby G. and Barnes J., p17, ASP Conf. Series volume 101
\bibitem[\protect\citeauthoryear{Band et al.}{1993}]{band_93} Band, D., Matteson, J., Ford, L., et al., 1993, ApJ 413, 281
\bibitem[\protect\citeauthoryear{Beardmore et al.}{2014}]{b7} Beardmore, A.P., Evans, P.A., Goad, M.R., et al., 2014, GCN \#17211
\bibitem[\protect\citeauthoryear{Bo\"er et al.}{2003}]{boer_05} Bo\"er, M., Klotz, A., Atteia, J.-L., et al., 2003, The Messenger 113, 45
\bibitem[\protect\citeauthoryear{Butler}{2007}]{but07} Butler, N.R., 2007, ApJ 656, 1001
\bibitem[\protect\citeauthoryear{Chevalier \& Li}{1999}]{che99} Chevalier, R.A., \& Li, Z.Y., 1999, ApJ 520, 29
\bibitem[\protect\citeauthoryear{Chevalier et al.}{2004}]{che04} Chevalier, R.A., Li, Z.Y., \& Fransson, C., 2004, ApJ 606, 369
\bibitem[\protect\citeauthoryear{Covino et al.}{2013}]{cov13} Covino, S., Melandri, A., Salvaterra, R., Campana, S., Vergani, S.~D., et al., 2013, MNRAS 432, 1231
\bibitem[\protect\citeauthoryear{Gehrels et al.}{2004}]{gehrels_04} Gehrels, N., Chincarini, G., Giommi, P., et al., 2004, ApJ 611, 1005
\bibitem[\protect\citeauthoryear{Gendre et al.}{2004}]{gen04} Gendre, B., Piro, L., \& De Pasquale, M., 2004, A\&A, 424, L27
\bibitem[\protect\citeauthoryear{Gendre et al.}{2006}]{gen07} Gendre, B., Corsi, A., \& Piro, L., 2006, A\&A 455, 803
\bibitem[\protect\citeauthoryear{Gendre et al.}{2008}]{gen08} Gendre, B., Galli, A., \& Bo\"er, M., 2008, ApJ 683, 620
\bibitem[\protect\citeauthoryear{Gendre et al.}{2012}]{gen12} Gendre, B., Atteia, J. L., Bo\"er, M., et al., 2012, ApJ 748, 59 
\bibitem[\protect\citeauthoryear{Gendre et al.}{2013}]{gen13} Gendre, B., Stratta, G., Atteia, J. L., et al., 2013, ApJ 766, 30 
\bibitem[\protect\citeauthoryear{Greiner et al.}{2008}]{greiner_08} Greiner, J., Bornemann, W., Clemens, Ch., et al., 2008, PAASP 120, 405
\bibitem[\protect\citeauthoryear{Greiner et al.}{2011}]{gre11} Greiner, J., Kr{\"u}hler, T., Klose, S., Afonso, P., Clemens, C., et al., 2011, A\&A 526, 30
\bibitem[\protect\citeauthoryear{Jakobsson et al.}{2004}]{jak04} Jakobsson, P., Hjorth, J., Fynbo, J.P.U., Watson, D., Pedersen, K., et al., 2004, ApJ, 617, L21
\bibitem[\protect\citeauthoryear{Kalberla et al.}{2005}]{kal05} Kalberla et al. 2005, A\&A, 440, 775
\bibitem[\protect\citeauthoryear{Kann et al.}{2006}]{kan06} Kann, D.A., Klose, S., Zeh, A., 2006, ApJ 641, 993,
\bibitem[\protect\citeauthoryear{Kann et al.}{2010}]{kan10} Kann, D.A., Klose, S., Zhang, B., Malesani, D., Nakar, E., et al., 2010, ApJ 720, 1513
\bibitem[\protect\citeauthoryear{Klotz et al.}{2008}]{klotz_08} Klotz, A., Vachier, F., Bo\"er, M., et al., 2008, AN 329, 275
\bibitem[\protect\citeauthoryear{Klotz et al.}{2014}]{b3} Klotz, A., Turpin, D., Bo\"er, M., et al., 2014, GCN \#17227
\bibitem[\protect\citeauthoryear{Kr{\"u}hler et al.}{2008}]{kru08} Kr{\"u}hler, T. K{\"u}pc{\"u} Yolda\c{s}, A., Greiner, J. Clemens, C. McBreen, S., et al., 2008, ApJ 685, 376
\bibitem[\protect\citeauthoryear{Kr{\"u}hler et al.}{2011}]{kru11} Kr{\"u}hler, T., Greiner, J., Schady, P., Savaglio, S., Afonso, P.M.J., et al., 2011, A\&A 534, 108
\bibitem[\protect\citeauthoryear{Kr{\"u}hler et al.}{2012}]{kru12} Kr{\"u}hler, T., Malesani, D., Milvang-Jensen, B., Fynbo, J.P.U., Hjorth, J., et al., 2012, ApJ 758, 46
\bibitem[\protect\citeauthoryear{Kumar \& Zhang}{2015}]{kum15} Kumar, P. \& Zhang, B., 2015, Physics Reports 561, 1
\bibitem[\protect\citeauthoryear{Marshall \& Sonbas}{2014}]{b11} Marshall, F.E. \& Sonbas, E., 2014, GCN \#17219
\bibitem[\protect\citeauthoryear{Maselli et al.}{2014}]{b10} Maselli, A., Melandri, A., Sbarufatti, B., et al., 2014, GCN \#17214
\bibitem[\protect\citeauthoryear{M\'esz\'aros \& Rees}{1997}]{mes97} M\'esz\'aros, P, \& Rees, M.J., 1997, ApJ, 476, 232
\bibitem[\protect\citeauthoryear{Meszaros}{2006}]{mes06} P. Meszaros, 2006, RPPh 69, 2259
\bibitem[\protect\citeauthoryear{Panaitescu et al.}{1998}]{pan98} Panaitescu, A., M\'esz\'aros, P., \& Rees, M.J., 1998, ApJ, 503, 314
\bibitem[\protect\citeauthoryear{Panaitescu \& Kumar}{2000}]{pan00} Panaitescu, A., \& Kumar, P., 2000, ApJ 543, 66
\bibitem[\protect\citeauthoryear{Pei}{1992}]{pei_92} Pei, Y.C., 1992, ApJ 395, 130
\bibitem[\protect\citeauthoryear{Perley et al.}{2014}]{b9} Perley, A.D., Cao, Y., Cenko, S.B., et al., 2014, GCN \#17228
\bibitem[\protect\citeauthoryear{Piro et al.}{2014}]{pir13} Piro, L., Troja, E., Gendre, B., et al., 2014, ApJ 790, 15
\bibitem[\protect\citeauthoryear{Racusin et al.}{2009}]{racusin_09} Racusin, J.L., Liang, E.W., Burrows, D.N., 2009, ApJ 698, 43
\bibitem[\protect\citeauthoryear{Rees \& M\'esz\'aros}{1992}]{ree92} Rees, M.J., \& M\'esz\'aros, P., 1992, MNRAS, 258, 41
\bibitem[\protect\citeauthoryear{Reichart et al.}{2005}]{rei05} Reichart, D., Nysewander, M., Moran, J, et al., 2005, Nuovo Cimento C, 28, 767
\bibitem[\protect\citeauthoryear{Rhoads}{1997}]{rho97} Rhoads, J.E., 1997, ApJ 487, L1
\bibitem[\protect\citeauthoryear{Rhoads}{1999}]{rho99} Rhoads, J.E., 1999, ApJ 525, 737
\bibitem[\protect\citeauthoryear{Romano et al.}{2006}]{romano_06} Romano, P., Campana, S., Chincarini, G., et al., 2006, A\&A 456, 917
\bibitem[\protect\citeauthoryear{Rossi et al.}{2012}]{ros12} Rossi, A., Klose, S., Ferrero, P., Greiner, J., Arnold, L.~A., et al., 2012, A\&A 545, 77
\bibitem[\protect\citeauthoryear{Sari \& Piran}{1999}]{sar99} Sari, R., and Tsvi, P., 1999, ApJ 520, 641
\bibitem[\protect\citeauthoryear{Schlafly \& Finkbeiner}{2011}]{schlafly_11} Schlafly, E.F. \& Finkbeiner D.P., 2011, ApJ 737, 103
\bibitem[\protect\citeauthoryear{Schweyer et al.}{2014}]{b8} Schweyer, T., Wiseman, P., Schady, P., et al., 2014, GCN \#17212
\bibitem[\protect\citeauthoryear{Sonbas et al.}{2014a}]{b1} Sonbas, E., Cummings, J. R., D'Elia, V., et al., 2014, GCN \#17206
\bibitem[\protect\citeauthoryear{Sparre \& Starling}{2013}]{spa12} Sparre, M., \& Starling, R.L.C., 2013, MNRAS 427, 2965
\bibitem[\protect\citeauthoryear{Starling et al.}{2013}]{rha12} Starling, R.L.C., Page, K.L., Pe'er, A., et al., 2013, MNRAS 427, 2950
\bibitem[\protect\citeauthoryear{Stratta et al.}{2004}]{str04} Stratta, G., Fiore, F., Antonelli, A., et al., 2004, ApJ, 608, 846
\bibitem[\protect\citeauthoryear{Trotter et al.}{2014a}]{b4} Trotter, A., Haislip, J., Reichart, D., et al., 2014, GCN \#17210
\bibitem[\protect\citeauthoryear{Trotter et al.}{2014b}]{b4b} Trotter, A., Haislip, J., Reichart, D., et al., 2014, GCN \#17221
\bibitem[\protect\citeauthoryear{Ukwatta et al.}{2014}]{b2} Ukwatta, T.N., Barthelmy, S.D., Baumgartner, W.H., et al., 2014, GCN \#17213
\bibitem[\protect\citeauthoryear{Vaughan et al.}{2005}]{vaughan_05} Vaughan, S., Goad, M.R., Beardmore, A.P., et al., 2005, ApJ 638, 920
\bibitem[\protect\citeauthoryear{Watson et al.}{2013}]{wat13} Watson, D., Zafar, T., Andersen, A.C., Fynbo, J.P.U., Gorosabel, J., et al., 2013,ApJ 768, 23
\bibitem[\protect\citeauthoryear{Woosley}{1993}]{woo93} Woosley, S.E., 1993, ApJ 405, 273
\bibitem[\protect\citeauthoryear{Yu H.F.}{2014}]{b12} Yu H.F., 2014, GCN \#17216
\bibitem[\protect\citeauthoryear{Zafar et al.}{2011}]{zaf11} Zafar, T., Watson, D., Fynbo, J.P.U., Malesani, D., Jakobsson, P., et al., 2011, A\&A 532, 143



\end{thebibliography}
\end{document}